\documentclass[preprint,floatfix,amsmath,amssymb,aps,prb]{revtex4-2}
\usepackage[T1]{fontenc}
\usepackage[utf8]{inputenc}
\usepackage{color}
\usepackage{verbatim}
\usepackage{amsmath}
\usepackage{amsthm}
\usepackage{amssymb}
\usepackage{graphicx}
\usepackage{esint}
\usepackage{dcolumn}
\usepackage{bm}
\usepackage{amsfonts}
\usepackage[table]{xcolor} %
\usepackage{subfigure}

\makeatletter

\usepackage{braket}

\makeatother
\usepackage{placeins}
\begin{document}

\renewcommand{\figurename}{Fig.}
\title{Spin-Exchange Induced Confinement of Correlated Bound States in the Continuum}
\author{Kai Chen$^{1,\dagger}$}
\author{Junyan Guan$^{1}$}
\author{Zhongming Gu$^{*}$}
\author{Jie Zhu$^{\dagger\dagger}$}
\affiliation{School of Physical Science and Engineering, Tongji University, Shanghai, China \\
$^{\dagger}$Corresponding author: KaiChenPhys@tongji.edu.cn \\
$^{*}$Corresponding author: zhmgu@tongji.edu.cn \\
$^{\dagger \dagger}$Corresponding author: jiezhu@tongji.edu.cn \\
$^{1}$These authors contribute equally to this work.}
\date{\today}
\begin{abstract}
Repulsively bound doublons are two-particle composites formed by strong interactions and are usually separated from the scattering continuum. Introducing spin-exchange coupling (SEC) fundamentally alters the underlying band structure, providing a powerful tuning knob to shift these isolated pairs toward this continuum. However, because entering such a regime typically dictates immediate dissociation, whether this coupling can drive these pairs inside while preserving their bound nature constitutes a fundamental unresolved challenge. Here we show that spin-exchange coupling in the one-dimensional extended Fermi–Hubbard model can drive doublons into the two-particle scattering continuum. Most of these states hybridize with extended channels and decay, whereas a subset remains decoupled and spatially bound, forming many-body bound states in the continuum (BICs). We map the interacting two-particle problem onto a two-dimensional lattice of coupled acoustic cavities, and experimentally observe both the radiating doublon continuum and the confined BIC states. These results demonstrate that spin-exchange coupling can turn selected doublons into interaction-induced BICs, deepening the understanding of continuum physics for interaction-bound pairs.
\end{abstract}

\maketitle
\raggedbottom
\setlength{\parskip}{0pt}

\section*{Introduction}
Repulsively bound doublons are essential for understanding how composite states can survive within or above a scattering continuum. Crucially, their stability often cannot be explained by spatial symmetries alone, as nearby extended states in the continuum may belong to the exact same symmetry sector. Instead, these states are protected by energy conservation: the exceptionally large on-site repulsion $U$ prevents a doublon from easily decaying into independent single-particle extended states \cite{winkler2006repulsively, strohmaier2010observation,hofmann2012doublon,xia2015quantum,chudnovskiy2012doublon,balzer2018doublon}. Because of this massive energy barrier, doublons are incredibly long-lived, boldly challenging the paradigm that interacting quantum systems must inevitably reach thermal equilibrium \cite{srednicki1994chaos, d2016quantum, deutsch2018eigenstate}. Consequently, their dynamics provide a unique probe into the limits of eigenstate thermalization, while also guiding the optimal initialization of polar molecules in optical lattices \cite{covey2016doublon}, sub-lattice localization \cite{bello2017sublattice}, and state preparation in near-term quantum devices \cite{paul2024realizing}.

Fundamentally, a doublon that remains perfectly localized while energetically embedded within a continuum of extended states is a direct manifestation of a BIC. BICs defy conventional scattering theory by completely decoupling from surrounding radiation channels, thereby possessing a theoretically infinite quality factor \cite{stillinger1975bound,plotnik2011experimental,zhen2014topological,hsu2016bound,marinica2008bound,chen2023observation}. This extraordinary confinement has driven immense interest across classical wave physics, where ultra-high-$Q$ quasi-BICs are utilized for extreme sound and light trapping, nonlinear enhancement, and precision filtering \cite{yin2025dimensional,guo2024realization,deriy2022bound,kang2023applications}. To overcome the inherent fragility of accidental BICs, recent advancements have focused on topological BICs, which leverage momentum-space topology to ensure robust protection against continuous structural perturbations \cite{zhen2014topological,hsu2016bound,xiao2017topological,cerjan2020observation,liu2023universal,hu2021nonlinear,bulgakov2017topological,benalcazar2020bound,qian2024non,qian2024non,bulgakov2017topological, chen2026layer}.

The connection between interaction-bound pairs and BICs motivates examining how a bound state remains confined when its energy overlaps extended states. Studies of interacting lattice systems have explored this coexistence \cite{huang2024interaction,zhao2026correlated,liu2024fate}, while acoustic implementations have made aspects of two-particle configuration space accessible to direct field measurements \cite{pu2026acoustic}. This connection also suggests a design strategy: use the internal configurations of a particle pair to introduce additional coupling pathways, then control their interference to preserve localization within a tunable continuum.

Here, inspired by the opposite-spin configurations of particle pairs in Hubbard-type models \cite{essler2005one}, we construct a bilayer acoustic lattice with engineered spin-exchange coupling. The two layers provide interfering hopping pathways, while the coupling $\lambda$ sets the effective hopping amplitudes $t_1-\lambda$ and $t_1+\lambda$ in the two layer-parity sectors. As $\lambda$ changes the continuum bandwidths, alternating-phase diagonal modes remain pinned at $E=U$ because their outgoing hopping amplitudes cancel. The continuum can therefore encompass these modes while their diagonal confinement is preserved. At $\lambda=t_1$, hopping cancellation extends throughout one parity sector, producing a flat band of localized states embedded in the opposite-sector continuum. We derive these confinement conditions and measure localized and extended responses at overlapping frequencies in a fabricated acoustic lattice. Our findings show how coupling between internal configurations controls spectral embedding and confinement, providing a route to acoustic BICs inspired by interacting-particle physics.

\begin{figure}[htbp]
\includegraphics[width=1\columnwidth]{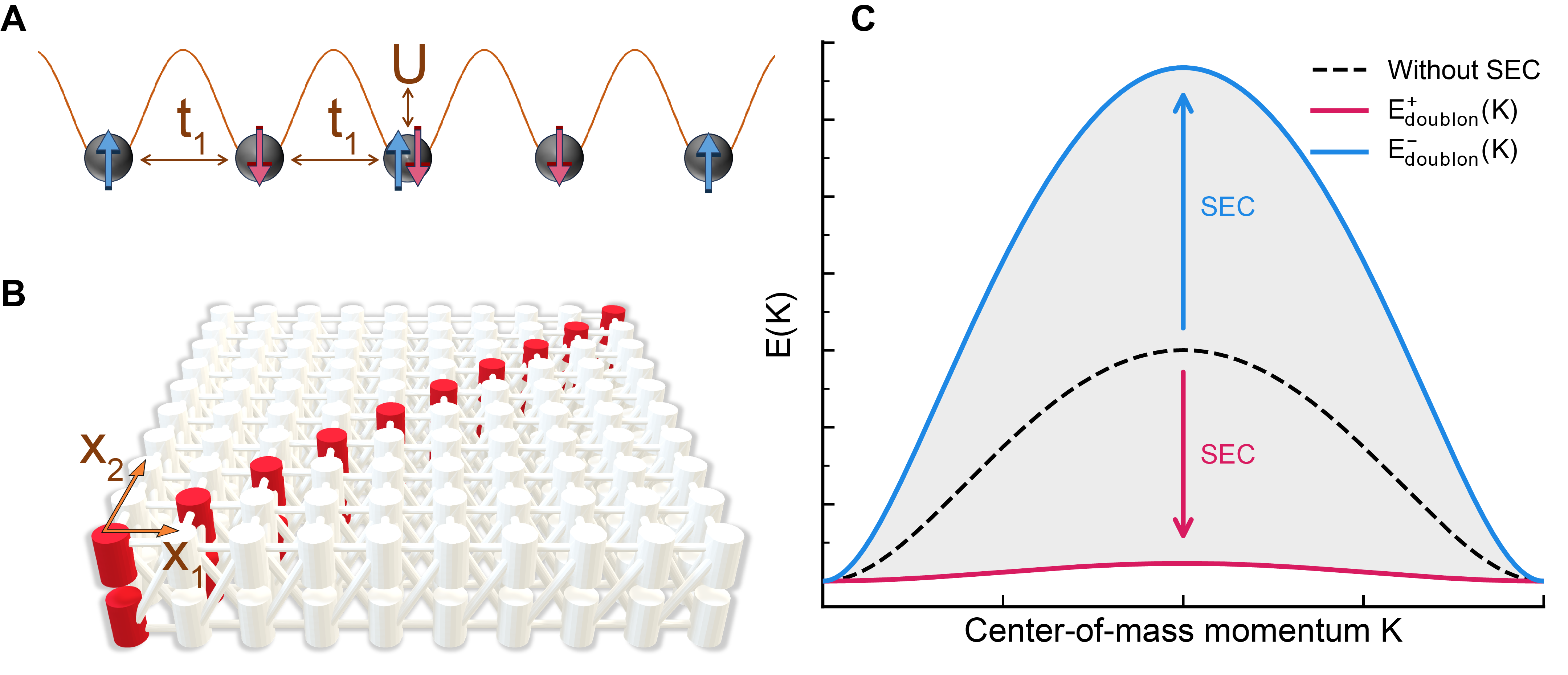} 
\caption{Bilayer lattice construction inspired by an interacting
particle model.
(A) Schematic of the one-dimensional extended Fermi--Hubbard model.
(B) Two-layer noninteracting lattice inspired by its two-particle
configuration-space description with SEC.
The on-site energy shift $U$ along the diagonals, marked by red
cylinders, represents the Hubbard interaction in the
configuration-space analogy.
(C) SEC-induced splitting of the doublon-like bands, with
eigenenergies given by Eq.~\ref{edb}.}
\label{fig0}
\end{figure}

\section*{Results}
\subsection*{Theoretical analysis} 
We consider two fermions in a one-dimensional Hubbard
model with a repulsive on-site interaction $U>0$ and spin-exchange
coupling. The Hamiltonian reads
\begin{align}
H ={}&
-t_1 \sum_j \sum_{\sigma\in\{\uparrow,\downarrow\}}
\left(
c_{j+1,\sigma}^{\dagger}c_{j,\sigma}
+\text{H.c.}
\right)
+U\sum_j n_{j,\uparrow}n_{j,\downarrow}
\nonumber\\
&-\lambda\sum_{j,\ell}\sum_{s=\pm1}
\left(
c_{j+s,\downarrow}^{\dagger}
c_{\ell,\uparrow}^{\dagger}
c_{\ell,\downarrow}
c_{j,\uparrow}
+\text{H.c.}
\right).
\label{eq:Hamiltonian}
\end{align}

Here, $c_{j,\sigma}^{\dagger}$ creates a fermion at site $j$
with spin $\sigma$, and
$n_{j,\sigma}=c_{j,\sigma}^{\dagger}c_{j,\sigma}$ is the
corresponding number operator.
The parameter $t_1$ is the nearest-neighbour hopping amplitude,
and the real parameter $\lambda$ is the spin-exchange coupling
strength.
The coupling exchanges the spins of the two fermions while
one of them hops to a neighbouring site.
The indices $j$ and $\ell$ run independently over all lattice
sites, so this process has the same amplitude at every
interparticle separation.
The sum includes only allowed nearest-neighbour steps $j\to j+s$,
and $\text{H.c.}$ denotes the Hermitian conjugate.

For a spinless 1D Hubbard model with two particles at coordinates $x_1$ and $x_2$, the system can be mapped onto a 2D single-particle square lattice in the $(x_1, x_2)$ plane with nearest-neighbour hopping amplitude $t_1$. The Hubbard interaction translates into an on-site energy $U$ for sites along the diagonal ($x_1=x_2$). Consequently, doublon states with energy $U$ correspond to modes localized on this diagonal.

However, in the spinful case, a general two-particle state is a superposition of four distinct spin configurations: $\lvert\uparrow\uparrow\rangle$, $\lvert\downarrow\downarrow\rangle$, $\lvert\uparrow\downarrow\rangle$, and $\lvert\downarrow\uparrow\rangle$. Due to the Pauli exclusion principle ($c_{j,\sigma}^\dagger c_{j,\sigma}^\dagger = 0$), fermions with parallel spins cannot occupy the same site. This structurally prohibits the $\lvert\uparrow\uparrow\rangle$ and $\lvert\downarrow\downarrow\rangle$ components from ever experiencing the on-site Hubbard interaction $U$. Therefore, in the strong correlation regime, the localized doublon is formed exclusively by the anti-parallel spin components that feel the $U$ potential. Based on these spin permutations, the 1D system can be mapped onto a two-layer 2D square lattice (Fig. \ref{fig0}A). Specifically, Layer A represents the configuration where particle 1 has spin-$\uparrow$ at $x_1$ and particle 2 has spin-$\downarrow$ at $x_2$, while Layer B represents the reversed configuration where particle 1 has spin-$\downarrow$ at $x_1$ and particle 2 has spin-$\uparrow$ at $x_2$. Within each individual layer, the Hamiltonian is identical to the spinless case, and the hopping between these two layers is determined entirely by the spin-exchange coupling term in the 1D system. The two-particle wavefunction in this truncated configuration space is:

\begin{equation}
|\Psi\rangle = \sum_{x_1, x_2} \left[ \psi_A(x_1, x_2) |x_1,\uparrow_1;x_2,\downarrow_2\rangle- \psi_B(x_1, x_2) |x_1,\downarrow_1;x_2,\uparrow_2\rangle \right],
\label{fock}
\end{equation}

where $\psi_A(x_1, x_2)$ and $\psi_B(x_1, x_2)$ represent the probability amplitudes in layer A and layer B, respectively. Substituting this state into the eigenvalue equation $H |\Psi\rangle = E |\Psi\rangle$ yields the following coupled equations (see Section I of the Supplemental Material for details):
\begin{equation}
\begin{aligned}
E \psi_A(x_1, x_2) &= -t_1 \sum_{s=\pm 1} \left[ \psi_A(x_1+s, x_2) + \psi_A(x_1, x_2+s) \right] + U \delta_{x_1, x_2} \psi_A(x_1, x_2) \\
&\quad + \lambda \sum_{s=\pm 1} \left[ \psi_B(x_1+s, x_2) + \psi_B(x_1, x_2+s) \right], \\
E \psi_B(x_1, x_2) &= -t_1 \sum_{s=\pm 1} \left[ \psi_B(x_1+s, x_2) + \psi_B(x_1, x_2+s) \right] + U \delta_{x_1, x_2} \psi_B(x_1, x_2) \\
&\quad + \lambda \sum_{s=\pm 1} \left[ \psi_A(x_1+s, x_2) + \psi_A(x_1, x_2+s) \right].
\end{aligned}
\label{H2d}
\end{equation}

\begin{figure}[htbp]
\includegraphics[width=1\columnwidth, height=0.5\textheight, keepaspectratio]{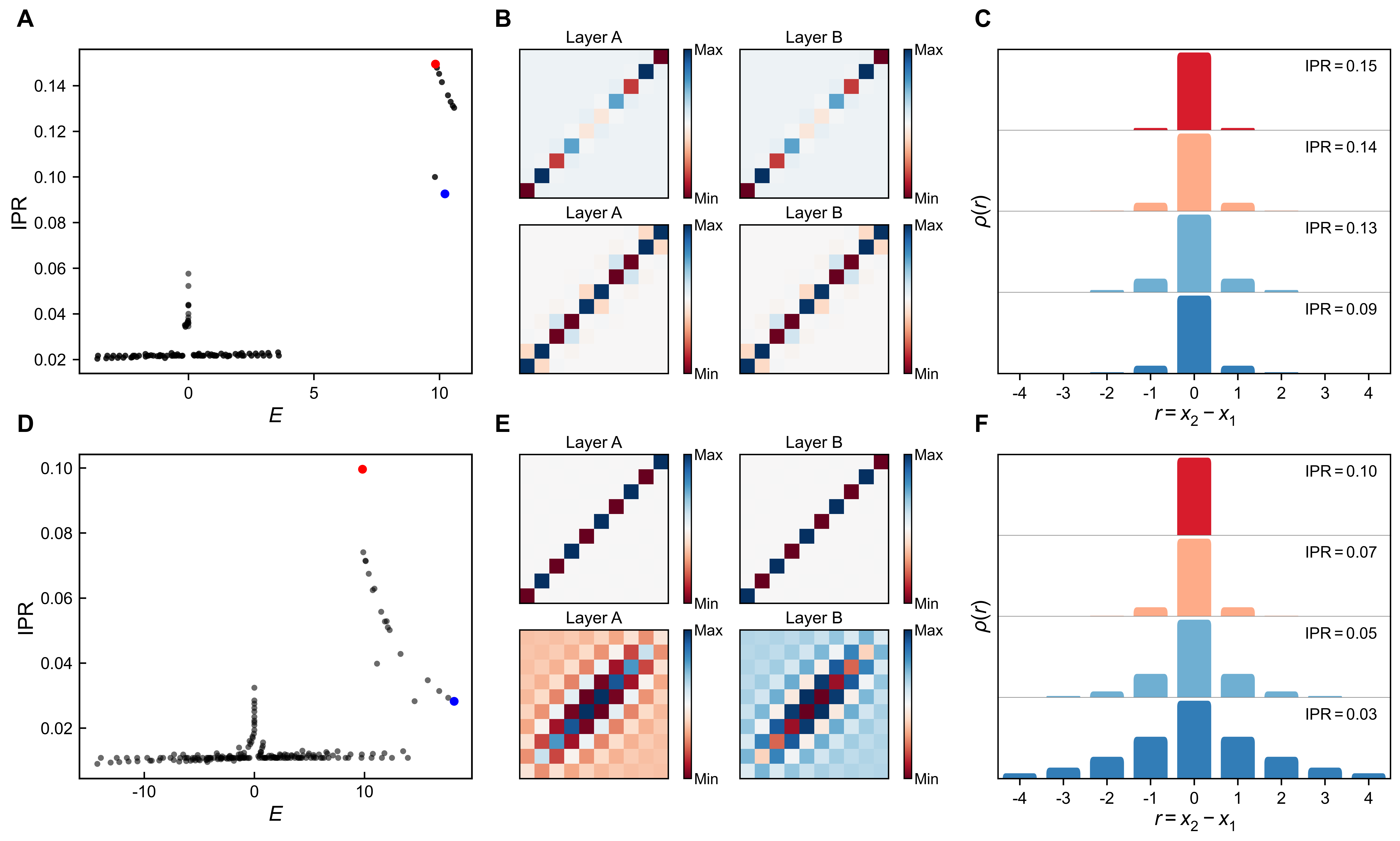}
\caption{Tight-binding calculations. (A) IPR versus eigenvalues in the absence of SEC. (B) Spatial distribution of typical doublon states. The upper panel displays the most localized state with the maximum IPR (red point in (A)), while the lower panel shows the highest-energy repulsive doublon state (blue point in (A)). (C) Diffusion function $\rho(r)$ with variable IPR in the absence of SEC. (D) IPR versus eigenvalues in the presence of SEC. (E) The upper panel shows the state with the maximum IPR (red point in (D)) and the state with the maximum energy eigenvalue (blue point in (D)). (F) Diffusion function $\rho(r)$ with variable IPR in the presence of SEC. The parameters are $t_1=1$, $\lambda=2.87$, and $U=9.82$.}
\label{fig1}
\end{figure}

Without spin-exchange coupling ($\lambda=0$), an interaction strength $U > 4|t_1|$ opens a spectral gap between the bound doublon states and the two-particle continuum (Fig.~\ref{fig1}A). For an $N\times N$ two-layer lattice, this yields $2N$ localized doublon states restricted to the spatial diagonals of each decoupled 2D layer. Although these states exhibit varying inverse participation ratios ($\text{IPR} \equiv \sum_n |\psi_n|^4 / (\sum_n |\psi_n|^2)^2$), they remain tightly confined to this diagonal (Fig.~\ref{fig1}B, C). 

Introducing spin-exchange coupling ($\lambda \neq 0$) breaks this isolation, splitting the single bound state into two distinct doublon bands (Fig. \ref{fig0}B) governed by the dispersion relation (see Section~II of the Supplemental Material):
\begin{equation}
E_{\text{doublon}}^\pm(K) = \sqrt{U^2 + 16(t_1 \mp \lambda)^2 \cos^2\left(\frac{K}{2}\right)},
\label{edb}
\end{equation}
where $K$ is the center-of-mass momentum of the doublon. Simultaneously, inter-layer symmetry becomes essential for accessing nontrivial phases~\cite{chen2026layer}. By exploiting the inter-layer swap symmetry $\mathcal{S}_l \equiv\sigma_x$, we block-diagonalize the Hamiltonian into symmetric and antisymmetric supermodes. This yields a continuum band dispersion $E_{\pm}(k_1, k_2) = -2(t_1 \mp \lambda)(\cos k_1 + \cos k_2)$, where $k_{1,2}$ are the single-particle Bloch momenta.

Consequently, for moderate interactions $U < 4 \max(|t_1 + \lambda|, |t_1 - \lambda|)$, the doublon energies are driven into this SEC-broadened continuum (Fig.~\ref{fig1}D). For generic doublon states, this spectral overlap induces scattering into the continuum, causing the stationary wavefunctions to spread into off-diagonal lattice sites (Fig.~\ref{fig1}E). To quantify this delocalization, we define the transverse spatial distribution $\rho(r) \equiv \sum_{x_2=x_1+r} \left( |\psi_A(x_1,x_2)| + |\psi_B(x_1,x_2)| \right)$. While $\rho(r)$ remains sharply peaked at $r=0$ for strictly confined states, it broadens significantly along the transverse $r$-direction for those hybridizing with the continuum (Fig.~\ref{fig1}F).

Remarkably, a specific subset of modes resists this hybridization. Protected by inversion symmetry, this state is characterized by the alternating-phase wavefunction $|\Psi_A\rangle = \frac{1}{\sqrt{N}} \sum_{x=0}^{N-1} (-1)^x |x, x\rangle_A$. The swap symmetry ($\mathcal{S}_l$) guarantees the existence of a perfectly degenerate partner state, $|\Psi_B\rangle$, in the opposite layer. The resulting symmetric and antisymmetric supermodes, $\left(|\Psi_A\rangle\pm|\Psi_B\rangle\right)/\sqrt{2}$, exhibit the maximum IPR in the system. Because these states are strictly confined to the diagonal with an alternating spatial phase, their kinetic hopping processes undergo perfectly destructive interference (see Section~III of the Supplemental Material). As a result, the only non-zero contribution to their energy arises from the on-site Hubbard interaction, yielding the exact eigenvalue relation $H |\Psi_{A/B}\rangle = U |\Psi_{A/B}\rangle$. Because this pinned energy level is entirely immune to the $\lambda$-induced broadening of the continuum bands, it emerges as a genuine BIC.

This symmetry protection reaches its ultimate extreme under the flat band condition $t_1=\lambda$. In this regime, the kinetic dispersion in the symmetric sector vanishes entirely ($E_+ = 0$). The model consequently hosts a macroscopic degeneracy of perfectly localized doublon states with a flat band energy of $U$, belonging exclusively to the symmetric sector of $\mathcal{S}_l$. For $0 < U < 8t_1$, the scattering continuum of the antisymmetric sector completely encompasses these states. Because they reside in strictly orthogonal symmetry sectors (see Section~IV of the Supplemental Material), this entire flat band of doublons is strictly protected from scattering, emerging as a massive family of symmetry-protected BICs.

\begin{figure}[htbp]
\includegraphics[width=1\columnwidth]{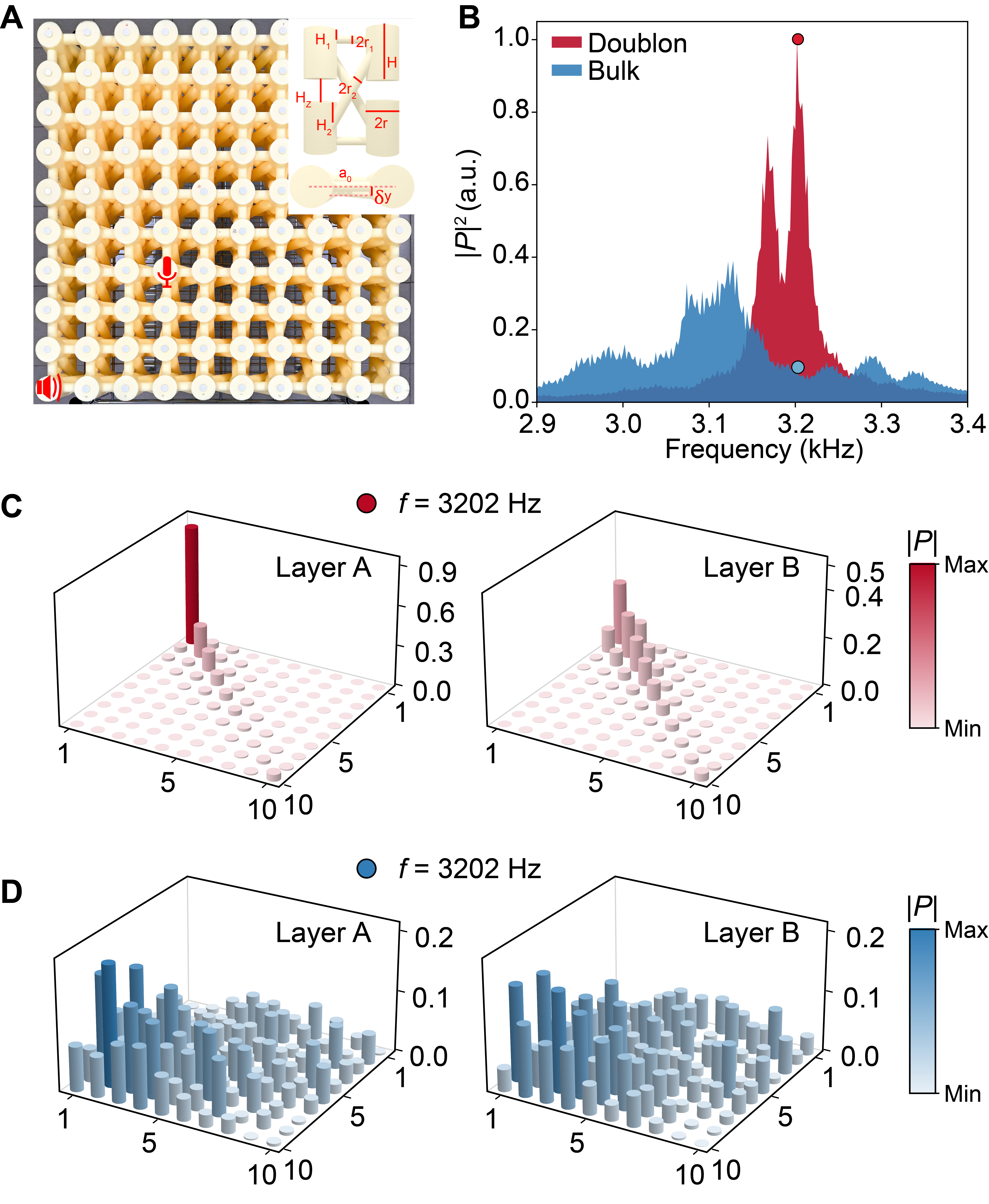}
\caption{Experimental setup, DOS, and typical pressure distributions. (A) The two-layer 2D acoustic lattice; the inset details the geometric parameters of a single unit cell. (B) DOS for the doublon modes (red) and the bulk states (blue). (C) Typical pressure distributions for the doublon mode indicated by the red point in (B). (D) Pressure distributions for the bulk modes at the frequency indicated by the blue point in (B).}
\label{fig2}
\end{figure}

\subsection*{Experimental results}
To experimentally confirm our theoretical predictions, we fabricated the corresponding acoustic lattice structures using 3D printing (Fig.~\ref{fig2}A). The baseline geometry is defined by a lattice constant $a_0 = 7.2\ \text{cm}$, total cavity height $H = 6\ \text{cm}$, main cavity radius $r = 1.92\ \text{cm}$, and intra-layer connecting channel radius $r_1 = 0.48\ \text{cm}$. The tight-binding parameters are mapped directly to these precise geometric features. Specifically, the intra-layer coupling strength ($t_1$) is tuned via the vertical offset of the connecting channels relative to the main cavity mid-height, given by $H/2 - H_1$ (where $H_1=H/3.6$). Similarly, the inter-layer coupling strength ($\lambda$) is controlled by an inter-layer channel with radius $r_2=0.54\ \text{cm}$, a vertical offset $H/2 - H_2$ ($H_2=0.72\ \text{cm}$), and a transverse shift $\delta y=0.76\ \text{cm}$ from the center of the cavity cross-section. The two layers are symmetrically joined across a mirror plane located at $z = H/2 + H_z/2$ (where $H_z=7.92\ \text{cm}$).

To synthetically implement the Hubbard $U$ term, which acts as an on-site energy in the 2D configuration space, we reduce the height of the cavities along the spatial diagonal of each layer by $\Delta H=0.38\ \text{cm}$ (indicated in red in Fig.~1A). Because this geometric perturbation inherently modifies the adjacent coupling strengths, we compensate by precisely tuning the channels connecting these diagonal cavities to their nearest neighbors. For these compensated sites, the intra-layer channel height is adjusted to $\tilde{H}_1=H/4.85$, while the inter-layer channels are modified to a radius of $\tilde{r}_2=0.66\ \text{cm}$ with a spatial shift of $\delta\tilde{y}=0.70\ \text{cm}$. With these calibrated parameters, the fabricated acoustic lattice achieves effective hopping ratios of $\lambda/t_1 \approx 2.87$ and $U/t_1 \approx 9.82$, successfully capturing the essential features of the tight-binding model.

To map these states experimentally, acoustic waves are injected using a localized single-point source (a loudspeaker). As denoted by the speaker icons in Fig.~\ref{fig2}A, this source is strategically positioned either on the diagonal of the lattice or in the bulk to selectively excite the doublon states or the bulk continuum modes, respectively. The steady-state acoustic pressure fields are obtained by measuring the pressure in each cavity of the acoustic crystal with a probe microphone. The measured density of states (DOS) in Fig.~\ref{fig2}B reveals the presence of doublon states embedded within the continuum bulk modes, demonstrating that the doublon state manifests as a BIC under non-zero spin-exchange coupling. The corresponding pressure-field distributions for a typical doublon state and a typical bulk state are shown in Figs.~\ref{fig2}C and \ref{fig2}D, respectively. 

\begin{figure}[htbp]
\includegraphics[width=1.\columnwidth]{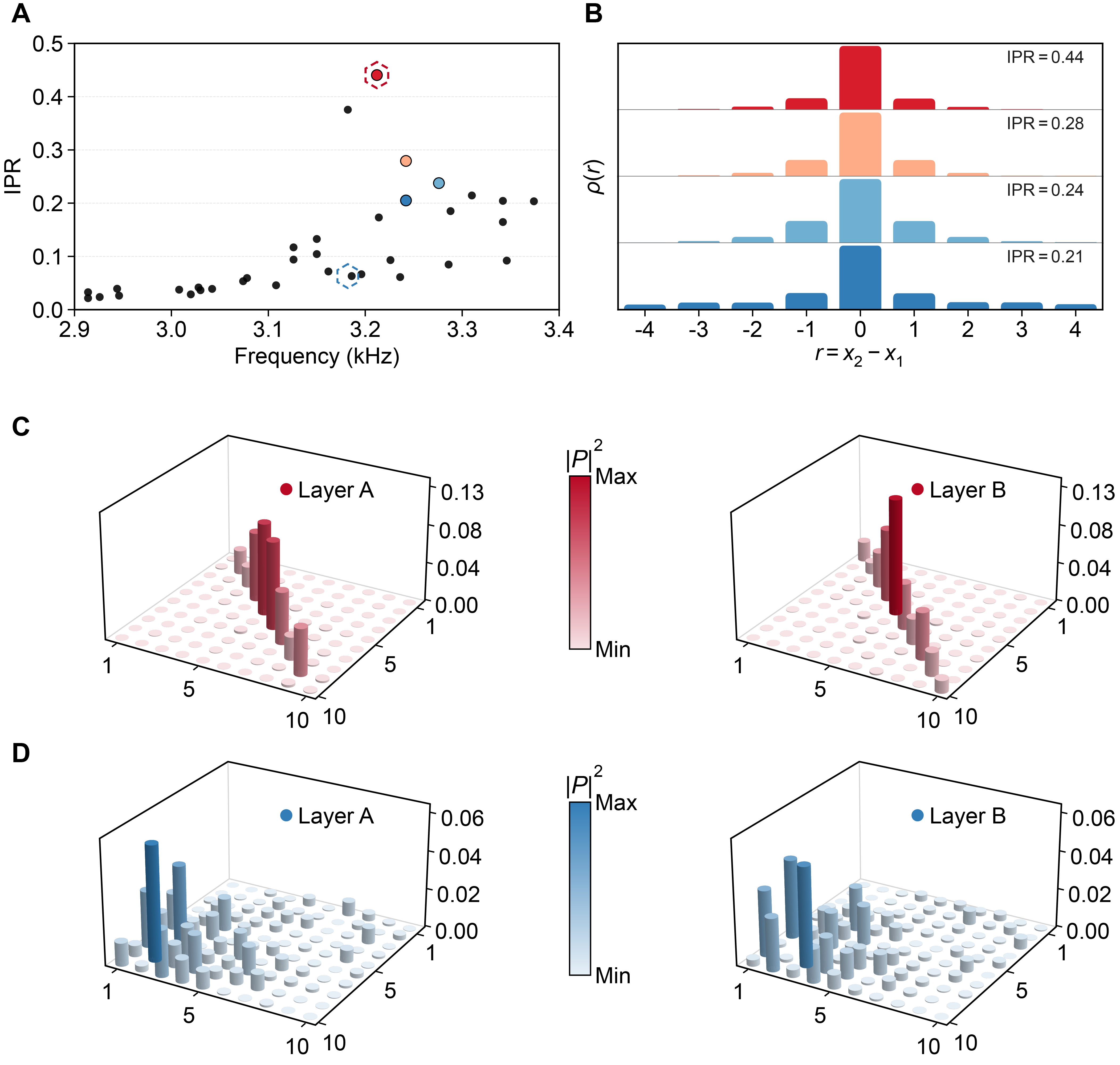}
\caption{Experimental measurements. (A) Measured IPRs. (B) Diffusion function $\rho(r)$ for modes with different IPRs. (C) Pressure distribution for the mode with the maximum IPR, indicated by the red dashed hexagon in (A). (D) Pressure distribution for the mode corresponding to the IPR marked by the blue dashed hexagon in (A).}
\label{fig3}
\end{figure}

Having established the existence of the doublon BICs, we probe the effects of spin-exchange coupling by measuring the inverse participation ratio, $\text{IPR} \equiv \sum_n |P_n|^4 / (\sum_n |P_n|^2)^2$, where $P_n$ is the acoustic pressure at the $n$-th site. As shown in Fig.~\ref{fig3}A, the measured IPR is in qualitative agreement with tight-binding calculations. Our theoretical model predicts that the doublon state with the maximum IPR exhibits an alternating phase along the diagonal while maintaining symmetric parity between the two layers. To selectively excite this mode and suppress boundary effects, we drive the system using 12 sources located along the diagonal (spanning sites 3 through 8 in each layer). Within each layer, these sources are driven out of phase (a $\pi$ phase shift) to enforce the alternating spatial profile, yet they remain strictly in phase across the two layers to isolate the symmetric supermode. The spatial distribution of this maximum-IPR doublon state (Fig.~\ref{fig3}C) reveals strong localization near the diagonals of each layer, confirming the perfectly destructive interference of kinetic hopping processes that isolates the BIC. Crucially, this doublon state is spectrally embedded within a continuum of lower-IPR bulk states (a typical bulk state is shown in Fig.~\ref{fig3}D), unambiguously demonstrating its nature as a BIC. Furthermore, the measured spatial diffusion function, $\rho(r)$ (Fig.~\ref{fig3}B), aligns with our theoretical analysis, confirming that spin-exchange coupling drives the diffusion of the remaining doublon states into the scattering continuum. Ultimately, our acoustic measurements directly validate the theoretical predictions, establishing that spin-exchange coupling can be leveraged to engineer BICs and manipulate nontrivial wave phenomena.

\section*{Discussion}
We have established that spin-exchange coupling in the 1D Hubbard model serves as a potent mechanism for driving doublon diffusion into the scattering continuum. Remarkably, however, by enforcing inter-layer swap symmetry, specific symmetric and antisymmetric supermodes are perfectly protected from this scattering, manifesting as robust BICs. These theoretical predictions are directly validated by our near-field acoustic measurements. Beyond the specific observation of these BICs, our methodology—mapping an interacting quantum many-body problem onto a highly tunable, non-interacting acoustic lattice—establishes a versatile experimental platform for complex wave physics. By demonstrating that synthetic spin-exchange coupling can actively dictate modal localization, our work opens fundamentally new avenues for engineering nontrivial wave states in both classical metamaterials and quantum synthetic matter.

\section*{Data availability}
The data that support the findings of this study are available from the corresponding author upon reasonable request.
\section*{Author contributions}
K.C. and J.Y.G. contributed equally to this work. K.C. developed the theoretical model, conducted the numerical simulations, and designed the acoustic lattice structure. J.Y.G. constructed the experimental setup, performed the measurements, and processed the experimental data. K.C., Z.M.G., and J.Z. supervised the research and drafted the manuscript. All authors contributed to the data analysis, discussed the physical implications, and reviewed and edited the final manuscript.

\section*{Competing interests}
The authors declare no competing interests.

\section*{Acknowledgements}
We acknowledge support from the National Key R\&D Program of China (Grants No. 2022YFA1404400 and No. 2022YFA1404403), the National Natural Science Foundation of China (Grants No. 92263208), the Fundamental Research Funds for the Central Universities. 



\bibliographystyle{apsrev4-2}
\bibliography{Lutlib}

\end{document}